\documentclass[twocolumn,prx,aps,superscriptaddress,longbibliography]{revtex4-2}
\usepackage{graphicx} 
\usepackage{amsmath}
\usepackage{amssymb}
\usepackage{siunitx}
    \DeclareSIUnit{\torr}{Torr}
\usepackage{hyperref} 
    \hypersetup{
        colorlinks=true,
        linkcolor=blue,
        urlcolor=blue,
        citecolor=blue}
\usepackage{float} 

\newcommand{\InO}{a:InO$_x$}

\begin{document}
\title{Excess noise in the anomalous metallic phase in amorphous indium oxide}
\author{Andr\'e Haug}
\email{andre-marcel.haug@weizmann.ac.il}
\affiliation{Department of Condensed Matter Physics, The Weizmann Institute of Science, Rehovot 76100, Israel}
\author{Dan Shahar}
\affiliation{Department of Condensed Matter Physics, The Weizmann Institute of Science, Rehovot 76100, Israel}
\date{\today}

\begin{abstract}
More than 25 years ago, unexpected metallic behavior was discovered on the superconducting side of the superconductor-to-insulator transition. To this day, the origin of this behavior is unclear. In this work, we present resistance and broadband voltage noise measurements in the kilohertz regime in amorphous indium oxide. We find that the metallic behavior gives rise to excess noise much larger than what is expected from thermal noise with an unexpected frequency and temperature dependence whose origin remains elusive.
\end{abstract}

\maketitle{}
Since the publication of the seminal paper by Abrahams et al. \cite{Abrahams.1979}, a metallic ground state was thought not to exist in two-dimensional (2D) systems at zero temperature ($T$). As a result, the ground state of 2D systems should be either superconducting or insulating. Some materials can be tuned from one ground state to the other, depending on parameters such as disorder, sample thickness, carrier concentration, or an externally applied magnetic field ($B$), coining the term superconductor-to-insulator transition (SIT) \cite{Haviland.1989,Jaeger.1989}.

In recent years, evidence for an unexpected metal-like phase on the superconducting side of the SIT has surfaced \cite{Ephron.1995,Mason.1999,Vaitiekėnas.2020,Tsen.2016,Breznay.2017,Yang.2019,Yang.2022,Saito.2015,Kravchenko.1994,Punnoose.2005}. The characteristics of this novel metallic state are (1) a temperature-independent plateau of the resistance ($R$) far below its value in the non-superconducting state and (2) a positive magnetoresistance of this resistance plateau. This unexpected metallic behavior has been termed the ``anomalous metal" (AM). Subsequent studies also found a vanishing Hall effect \cite{Breznay.2017} and the lack of cyclotron resonance \cite{Wang.2018}, providing evidence for particle-hole symmetry and short-range superconducting correlations, respectively. Despite decades of study, the underlying physics of the AM is poorly understood. Some argue that the metallic state is bosonic \cite{Das.2001,Das.1999,Dalidovich.2002,Phillips.2003,Tsen.2016}, but other mechanisms such as vortex tunneling causing the metallic behavior have been proposed as well \cite{Saito.2015,Ephron.1995}.

More recently, Tamir et al. \cite{Tamir.2019.02} eliminated the metallic behavior by implementing low-pass filters into their measurement leads, reducing the high-frequency radiation reaching the sample. As a result, the resistance decreased monotonously with temperature down to the noise floor of their measurement. Additionally, they showed that the saturation was re-introduced through Joule heating of the electrons by increasing the measurement bias current. They therefore concluded that the metallic behavior is a result of electron overheating from the high-frequency radiation rather than a novel phase of matter. Despite the use of filters in subsequent studies \cite{Dutta.2019,Dutta.2022,Doron.2020}, it has not always been possible to eliminate the AM \cite{Dutta.2022}.

Electron overheating is a well-known issue in 2D systems \cite{Doron.2020,Levinson.2016}. At low temperatures, electron overheating can be caused by a thermal weak-link between electrons and phonons. This limits the maximum power that the electrons can dissipate through the phonons. High-frequency radiation that couples directly to the electrons stemming from e.\,g. electronic devices or radio stations can exceed this power. As a result, the weak coupling of electrons and phonons at low temperature causes the electrons to remain at an elevated temperature $T_\mathrm{el}$ that is higher than the phonon temperature $T_\mathrm{ph}$.

In most experiments, we use resistance thermometers that must be electrically insulated from the sample to measure the temperature. Therefore, they are only thermally coupled to the sample via the phonons and hence measure $T_\mathrm{ph}$. In order to ensure equilibrium conditions, resistance measurements are repeated at various currents and thus various electrical powers. If the resistance does not change with the bias current, it is then assumed that the sample is at thermal equilibrium and $T_\mathrm{el}=T_\mathrm{ph}$.

However, if the power that stems from the bias current is small compared to the power absorbed by the electrons from high-frequency radiation, electron overheating can be caused by the latter rather than the former. Therefore, varying the high-frequency radiation by implementing low-pass filters with varying cutoff frequencies should be performed to ensure $T_\mathrm{el}=T_\mathrm{ph}$ while only varying the bias current leads to assuming $T_\mathrm{el}=T_\mathrm{ph}$ erroneously. Further cooling of the sample then decreases $T_\mathrm{ph}$ while $T_\mathrm{el}$ remains constant. As a result, the resistance does not change anymore with $T_\mathrm{ph}$ \cite{Tamir.2019.02}.
\begin{figure*}
    \begin{minipage}[c]{.49\textwidth}
        \begin{flushleft}
            \textbf{a}
        \end{flushleft}
        \centering
        \includegraphics[width=\textwidth]{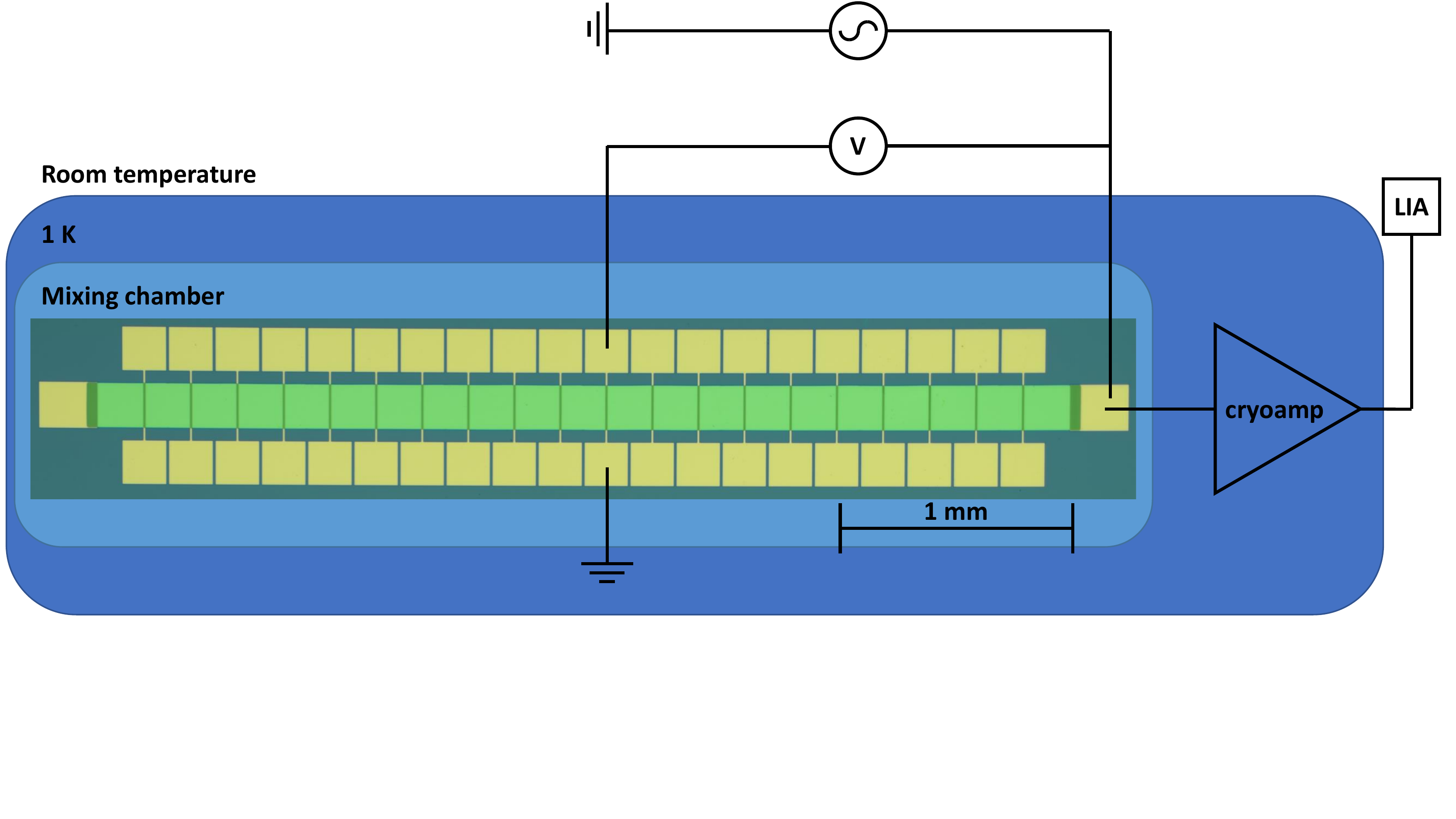}
    \end{minipage}
    \begin{minipage}[c]{.49\textwidth}
        \centering
        \includegraphics[width=\textwidth]{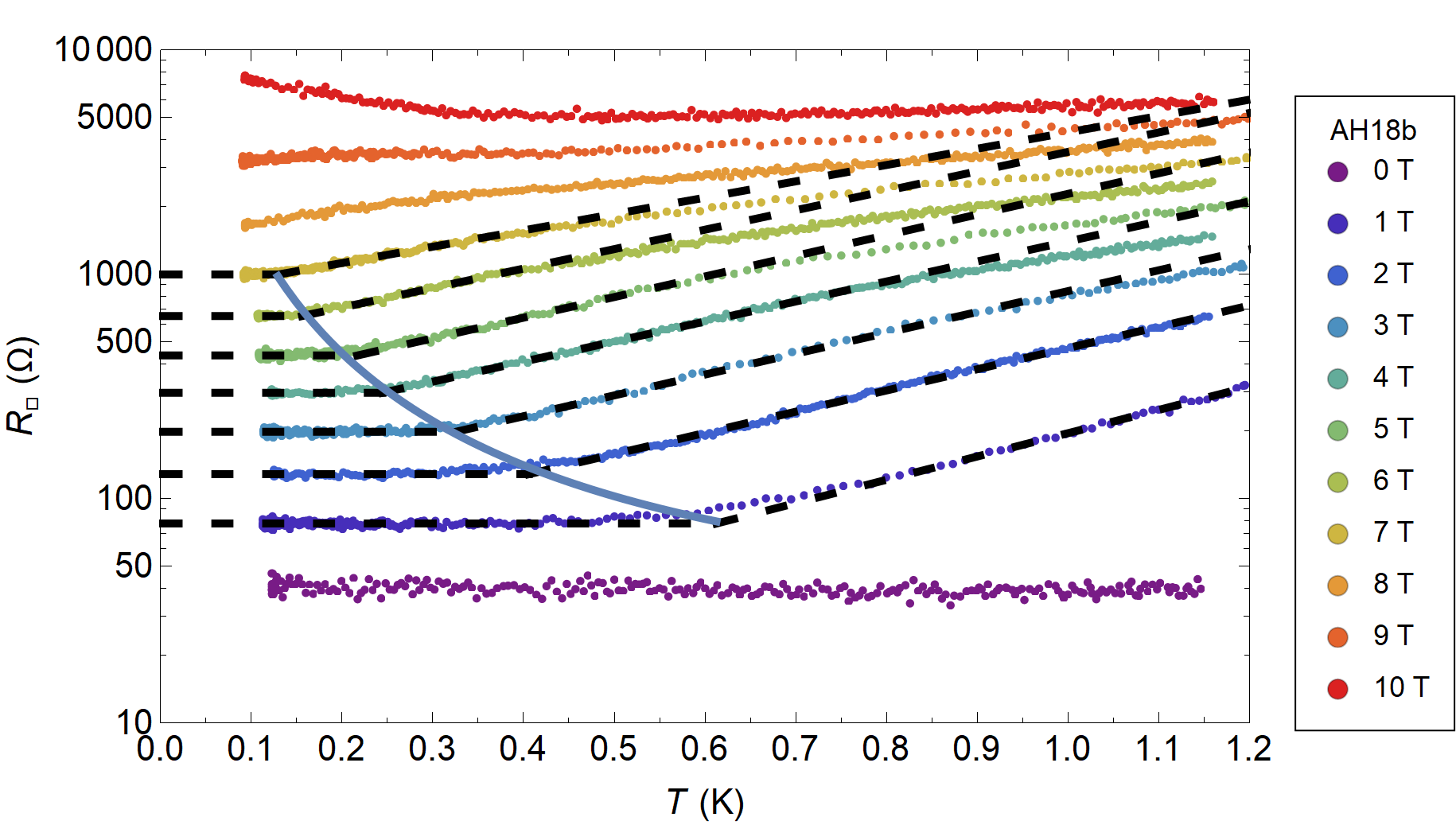}
    \end{minipage}
    \caption{\textbf{a} Experimental setup. A LIA both serves as current source and measures the voltage drop across the sample is used for the resistance measurement. The cryoamp and the LIA on the right are used for the noise measurement. The high number of contacts allows for a separation of the physics that are governed by the sheet resistance from the noise which only depends on the total resistance. For this experiment, we measured 10 squares. \textbf{b} Sheet resistance as a function of temperature at various magnetic fields of sample AH18. The black dashed lines indicate constant fits at low temperature and exponential fits at high temperature. The solid blue line is a guide to the eye for the transition temperature $T^\ast$.}
    \label{fig:exp_setup_resistance}
\end{figure*}

The purpose of the experiment described on these pages was to test the hypothesis of Tamir et al. that the metallic behavior is a result of electron overheating. We designed and conducted an experiment to measure $T_\mathrm{el}$ directly by means of broadband Johnson noise thermometry \cite{Johnson.1928,Nyquist.1928} in amorphous indium oxide (\InO). Johnson noise results from a time-dependent, random potential difference due to thermal fluctuations of charge carriers. At thermodynamic equilibrium, its mean-square value per unit bandwidth ($S_\mathrm{J}$) is given by
\begin{equation}
    S_\mathrm{J} = \frac{V^2}{\Delta f} = 4 k_\mathrm{B} R T_\mathrm{el},
    \label{eqn:johnson_noise}
\end{equation}
where $k_\mathrm{B}$ is the Boltzmann constant and $\Delta f$ is the bandwidth over which the signal is measured. $T_\mathrm{el}$ can be directly obtained from Equation \eqref{eqn:johnson_noise} after measuring both $S_\mathrm{J}$ and $R$. Unlike commonly used resistance thermometers, which are secondary thermometers and have to be calibrated, Johnson noise thermometry is a primary type of thermometry and therefore requires no calibration \cite{Ott.1988}.

Our main finding is a noise level much greater than what is to be expected from Johnson noise inside the AM. This excess noise increases with frequency and is proportional to $T_\mathrm{ph}^{-2}$ down to our base temperature of $T_\mathrm{ph} = \SI{140}{\milli\kelvin}$. Both the frequency and temperature dependence are surprising because Johnson noise is expected to decrease with temperature and to be frequency independent within the range of our experiment. This is indicative of an additional noise component that appears, in our experiments, only in AM regime.

Our \InO\ sample was grown by electron beam evaporation of In$_2$O$_3$ ingots in an oxygen-enriched environment at a pressure of $\SI{4.4E-6}{\torr}$ onto silicon substrate with a silicon dioxide layer. The film thickness was controlled during growth using a crystal monitor that was calibrated by measuring the film thickness of prior samples using atomic force microscopy.

All measurements were performed using a top-loading dilution refrigerator. We measured the resistance in a 3-probe rather than the standard 4-probe configuration in order to increase the bandwidth of the noise measurement (see Supplementary Material). Temperature was varied from $\SI{120}{\milli\kelvin}$ to $\SI{1.2}{\kelvin}$ at magnetic fields from $\SI{0}{\tesla}$ to $\SI{10}{\tesla}$. The resistance was measured using a lock-in technique with a frequency of $f=\SI{3.11}{\hertz}$; noise measurements were performed using a Stahl Electronics CX-4 cryogenic amplifier (cryoamp) and a Zurich Instruments HF2LI lock-in amplifier (LIA). The resistance measurement setup was disconnected at room temperature every time the noise was measured to eliminate unwanted noise. A schematic of the measurement setup is shown in Figure \ref{fig:exp_setup_resistance}a. We measured the voltage noise at different temperatures, tuning the system in and out of the AM while resistance was kept constant at $R=\SI{1.29}{\kilo\ohm}$ and $R=\SI{3.00}{\kilo\ohm}$, respectively, by adjusting the magnetic field accordingly. Measuring the noise at a constant resistance allows us to attribute all changes in the noise to the sample instead of e.\,g. the cryoamp's input current noise $i_\mathrm{n}$ dropping an additional $R$-dependent voltage $i_\mathrm{n} R$ across the sample.

In Figure \ref{fig:exp_setup_resistance}b, we show the sheet resistance ($R_\square$) as a function of temperature for magnetic fields from $\SI{0}{\tesla}$ to $\SI{10}{\tesla}$. Even at $B=0$, the resistance does not drop to zero. We attribute this residual resistance to our 3-probe configuration. As $B$ is increased, the signature of the AM can be seen from $\SI{1}{\tesla}$ to $\SI{7}{\tesla}$. At low temperatures, the resistance is independent of temperature. Above the transition into the AM, the resistance increases exponentially with temperature and shows the well-known thermally activated behavior \cite{Ephron.1995}. We extrapolated both the constant resistance at low temperature and the exponential behavior at high temperature and define the intersections of these extrapolations as the transition temperature $T^\ast$ at which the system enters the AM, with $T^\ast$ as high as $\SI{613}{\milli\kelvin}$ at $B = \SI{1}{\tesla}$. We find that, as expected from previous studies \cite{Ephron.1995,Tamir.2019.02}, $T^\ast$ decreases as $B$ is increased. At $\SI{10}{\tesla}$, the slope of the resistance changes sign and the sample becomes more insulating as $T \to 0$ as expected for the SIT.
\begin{figure*}
    \begin{minipage}[c]{.49\textwidth}
        \begin{flushleft}
            \textbf{a}
        \end{flushleft}
        \centering
        \includegraphics[width=\textwidth]{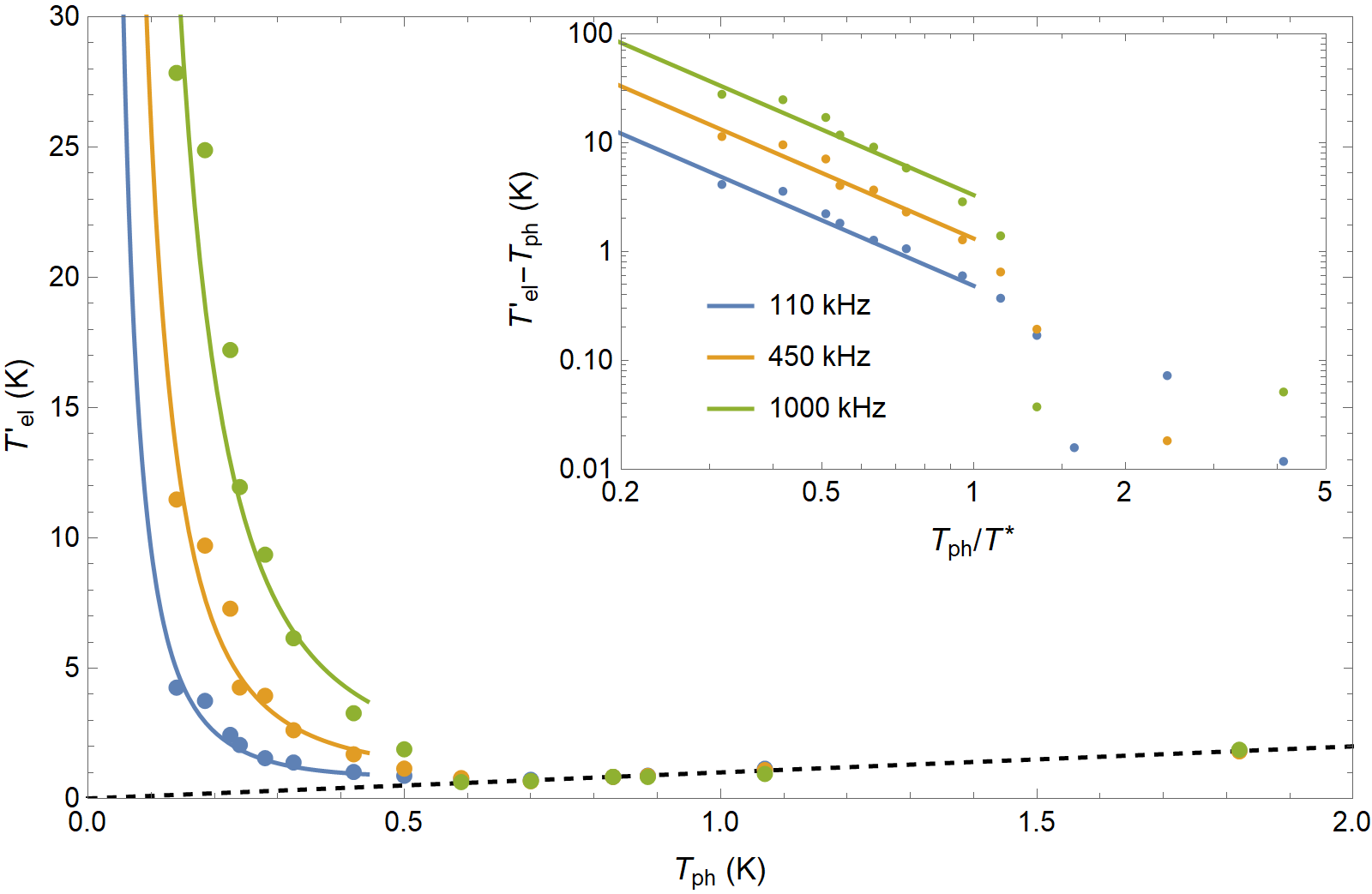}
    \end{minipage}
    \begin{minipage}[c]{.49\textwidth}
        \begin{flushleft}
            \textbf{b}
        \end{flushleft}
        \centering
        \includegraphics[width=\textwidth]{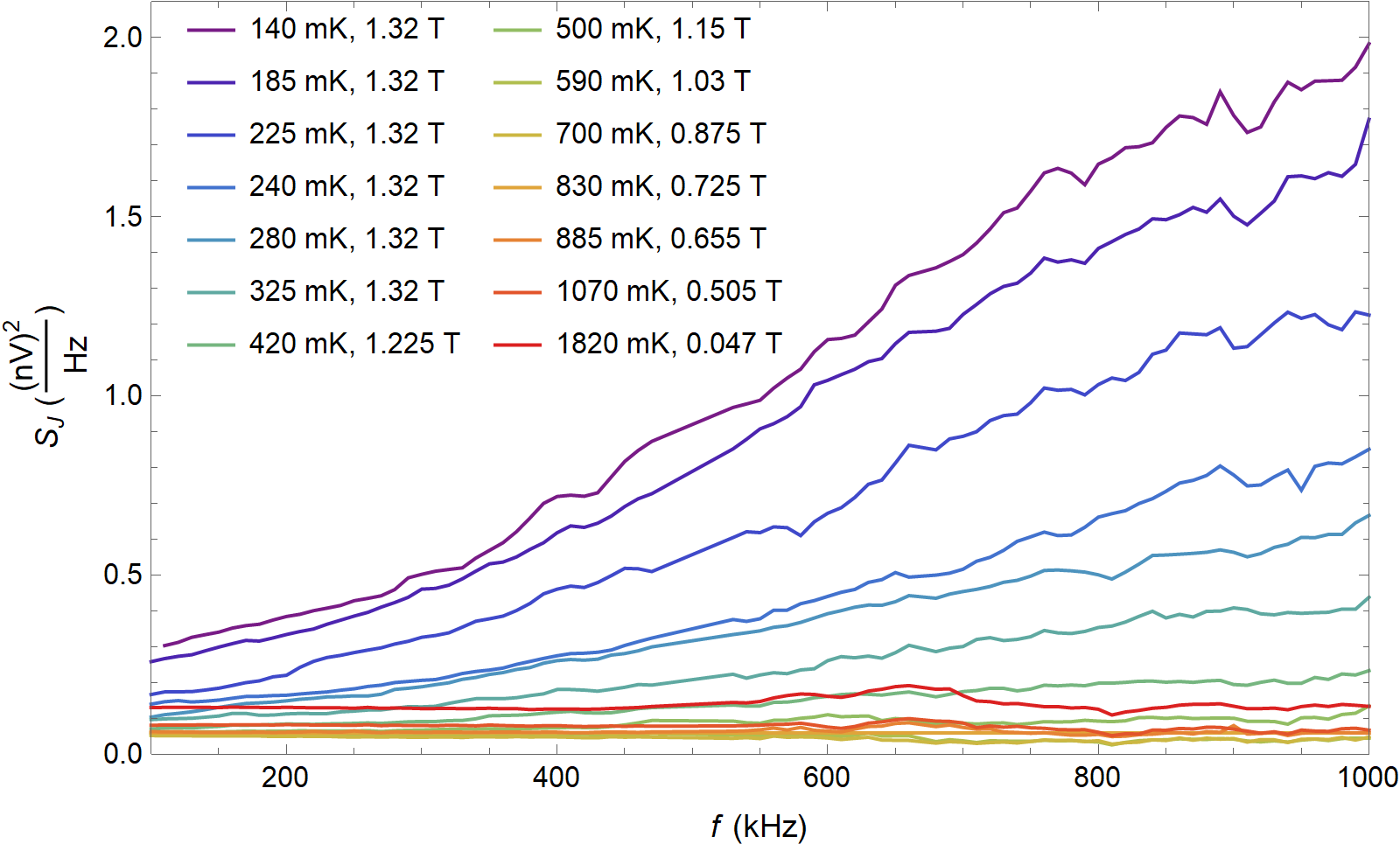}
    \end{minipage}
    \caption{\textbf{a} \textbf{Main panel} Ostensible electron temperature $T_\mathrm{el}^\prime = S_\mathrm{J}/4k_\mathrm{B}R$ as a function of $T_\mathrm{ph}$ for $R=\SI{1.29}{\kilo\ohm}$. The dashed line marks $T_\mathrm{el}^\prime = T_\mathrm{ph}$. For $T > T^\ast = \SI{438}{\milli\kelvin}$, $T_\mathrm{el}^\prime$ and $T_\mathrm{ph}$ are in good agreement. We therefore identify it as the true electron temperature $T_\mathrm{el}$ in this regime. At lower temperature, $T_\mathrm{el}^\prime$ starts to rise, assuming that all noise we measure is Johnson noise. For $T_\mathrm{ph} < T^\ast$, $T_\mathrm{el}^\prime$ appears to increase with $T_\mathrm{ph}^{-2}$. Solid lines are fits $\propto T_\mathrm{ph}^{-2}$. \textbf{Inset} Difference of $T_\mathrm{el}^\prime$ and $T_\mathrm{ph}$ as a function of $T_\mathrm{ph}$ renormalized by $T^\ast$. As $T_\mathrm{ph}/T^\ast$ approaches unity, $T_\mathrm{el}^\prime$ starts to increase. \textbf{b} Noise after accounting for amplifier noise for $R=\SI{1.29}{\kilo\ohm}$. At $T_\mathrm{ph} > T^\ast$, the noise is frequency-independent as is expected for Johnson noise. At $T_\mathrm{ph} < T^\ast$, the noise starts to increase and develops a frequency-dependence.}
    \label{fig:results}
\end{figure*}

After establishing the occurrence of the AM in our sample, we turn to the noise measurement. We measured the voltage noise at a constant resistance of $R=\SI{1.29}{\kilo\ohm}$ and $R=\SI{3.00}{\kilo\ohm}$ (see Supplemental Material for details) at various temperatures and magnetic fields over a frequency range from $\SI{100}{\kilo\hertz}$ to $\SI{1}{\mega\hertz}$. For reference, we use data above $\SI{800}{\milli\kelvin}$, where we assume $T_\mathrm{el}=T_\mathrm{ph}$, enabling us to distinguish between noise caused by the sample ($S_\mathrm{J}$) and other noise sources. The results for $R = \SI{3.00}{\kilo\ohm}$ can be found in the Supplemental Material. The results for $R=\SI{1.29}{\kilo\ohm}$ are shown in Figure \ref{fig:results}. In Figure \ref{fig:results}a, we show the Johnson noise term as a function of $T_\mathrm{ph}$. For $T_\mathrm{ph} > T^\ast$, we interpret this noise to stem from the true electron temperature $T_\mathrm{el}$. In this regime, $T_\mathrm{el}$ and $T_\mathrm{ph}$ agree within our error of $\SI{15}{\percent}$. For $T_\mathrm{ph} < T^\ast$, we observe an increase of the noise down to the lowest $T_\mathrm{ph}$ of $\SI{140}{\milli\kelvin}$ that is proportional to $T_\mathrm{ph}^{-2}$. Concomitantly, a frequency dependence starts to develop, with higher noise at higher frequencies. The frequency-dependent noise after accounting for amplifier noise is shown in Figure \ref{fig:results}b. We therefore decided to label this temperature as the ostensible electron temperature $T_\mathrm{el}^\prime$ since there is no physical explanation for the true electron temperature to increase as $T_\mathrm{ph}$ is decreased.

This behavior raises two issues: Firstly, Johnson noise is white noise and therefore should not exhibit any frequency dependence. Secondly, the noise appears to diverge as $T \to 0 $ which violates conservation of energy.

We are not aware of a scenario in which the noise increases as $T_\mathrm{ph}$ is decreased. A possible explanation for the behavior of the noise lies in the way we conduct our analysis: After calibration, we assume that (1) all noise is Johnson noise and (2) the electrons are at equilibrium. If one of these assumptions does not hold at $T_\mathrm{ph} < T^\ast$, the conversion from noise to temperature is no longer possible without appropriate theory.

Excess noise has already been found on the insulating side of the SIT \cite{Tamir.2019.01}, where it was interpreted in the context of many-body localization \cite{Basko.2007}. However, there has been no experimental evidence for excess noise on the superconducting side up until now and a theoretical prediction is still lacking.

We want to address two possible mechanisms for the behavior of the noise. The first one is kinetic inductance ($L_\mathrm{K}$), which arises from the inertia of the charge carriers. If large enough at low temperature, $L_\mathrm{K}$ would transform our setup from Figure \ref{fig:exp_setup_resistance}a into an RLC circuit with a resonance frequency of $f_\mathrm{r} = 1/(2\pi\sqrt{LC})$. Given our system's capacitance of approximately $C = \SI{500}{\pico\farad}$, the inductance would have to be roughly $L = \SI{2}{\micro\henry}$ for a resonance frequency of $f_\mathrm{r} = \SI{5}{\mega\hertz}$. Based on previous ac transport measurements \cite{Crane.2007}, we estimate $L_{\mathrm{K}\square} = \frac{m_\mathrm{e}}{n_\mathrm{s} e^2} = \SI{1.18E-16}{\henry}$ \cite{Lobb.1983}, corresponding to a resonance frequency of $f_\mathrm{r} \approx \SI{207}{\giga\hertz}$. We therefore discard kinetic inductance as the source of the evolving frequency dependence.

Moreover, $L_\mathrm{K}$ would only explain the frequency dependence of the noise, but not the excess noise that goes beyond the contribution of Johnson noise alone and is therefore not a suitable explanation for the unexpected behavior of the noise.

The second mechanism we would like to address is ergodicity \cite{Neumann.1929}. A system is ergodic when all states that are available to a system are being occupied at equal probabilities. Ergodicity breaking has been tied to many-body localization theoretically \cite{Luca.2013}. If the electrons of our sample behave in a non-ergodic way, the assumption that they are equilibrated with one another breaks down we cannot infer $T_\mathrm{el}$ from our noise measurement.  This could lead to the resistance being dominated by the cold electrons, whereas the hot electrons might be responsible for the increase in noise. Unfortunately, while this approach does explain the excess noise we see in the AM, it does not make a statement about the unexpected frequency dependence.

In conclusion, we found excess noise in the AM in \InO. This excess noise increases with frequency and appears to diverge $\propto T_\mathrm{ph}^{-2}$ as $T_\mathrm{ph}$ decreases. Both these features are unexpected since the noise must vanish as $T \to 0$ and $f \to \infty$ to conserve energy. We discussed two possible origins of this excess noise, namely kinetic inductance and ergodicity breaking, but no conclusions can be drawn at this early stage. More experimental evidence is needed at low temperatures and higher frequencies to tie the excess noise to the AM.
\section*{Acknowledgements}
We are grateful to P. Armitage, M. Feigel'man, and B. Sac\'ep\'e for fruitful discussions. This research was supported by the United States -- Israel Binational Science Foundation (BSF Grant No. 2012210).

\bibliography{bibliography}

\newpage
\onecolumngrid
\begin{center}
    \textbf{\large Excess noise in the anomalous metallic phase in amorphous indium oxide}\\\vspace{0.2cm}
    \textbf{\small SUPPLEMENTAL MATERIAL}\\\vspace{0.3cm}
     Andr\'e  Haug and Dan Shahar
\end{center}

\renewcommand{\thepage}{S\arabic{page}}
\renewcommand{\thefigure}{S\arabic{figure}}
\renewcommand{\thetable}{S\arabic{table}}
\renewcommand{\theequation}{S\arabic{equation}}
\renewcommand{\thesection}{S\arabic{section}}

\setcounter{page}{1}
\setcounter{figure}{0}
\setcounter{table}{0}
\setcounter{equation}{0}
\setcounter{section}{0}

\twocolumngrid
\section{Amplifier calibration}
\begin{figure}
    \centering
    \includegraphics[width=\linewidth]{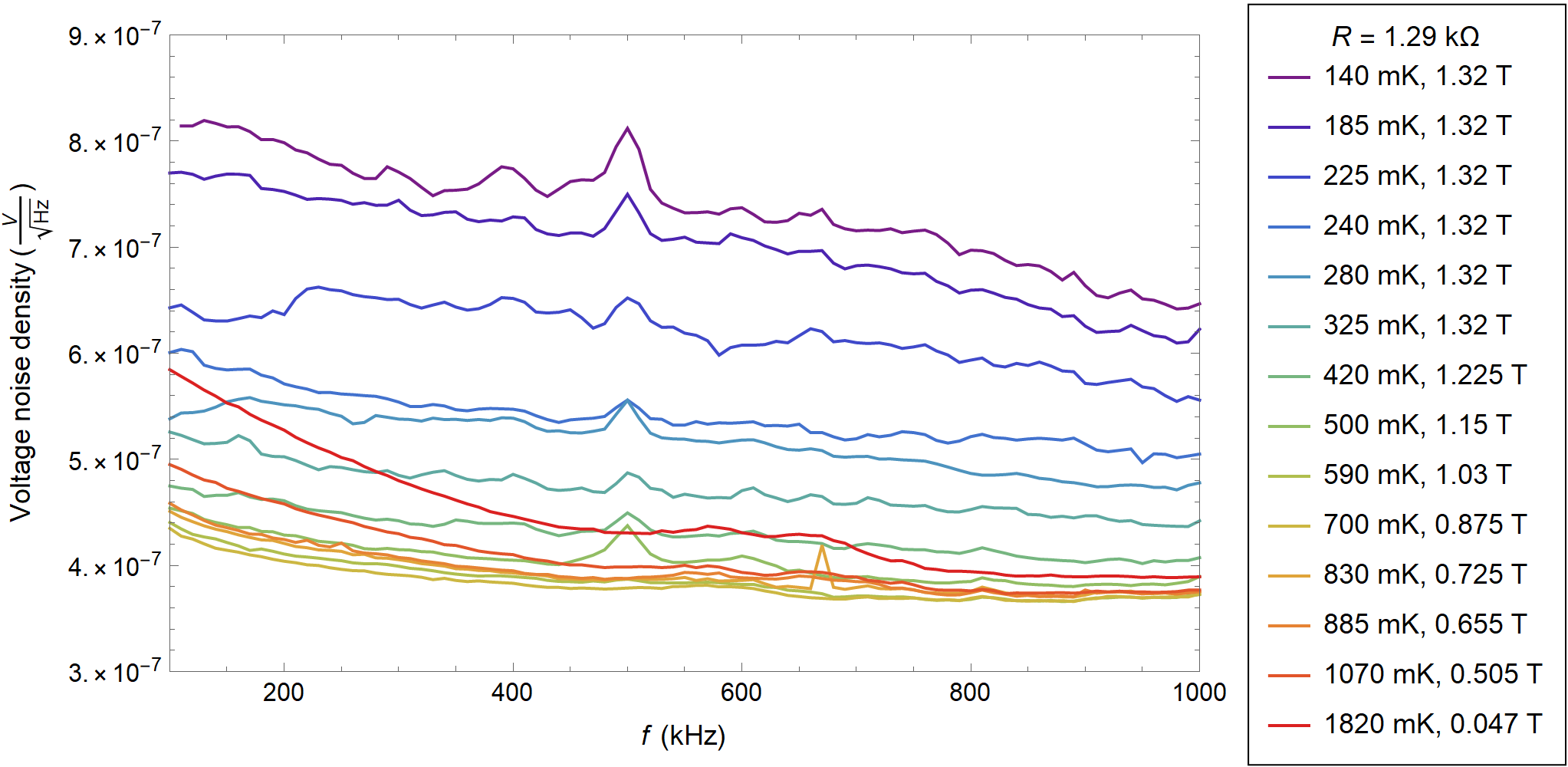}
    \caption{Voltage noise density as measured with our HF2LI LIA for $R = \SI{1.29}{\kilo\ohm}$. Data have been taken at 91 equidistant frequencies from $\SI{100}{\kilo\hertz}$ to $\SI{1}{\mega\hertz}$ with $\num{5E6}$ samples per frequency, a fourth-order low-pass filter and a time constant of $T_\mathrm{c}=\SI{7.812}{\micro\second}$ (BW NEP $\Delta f = \SI{10}{\kilo\hertz}$).}
    \label{supp_fig:raw_noise_1kOhm}
\end{figure}

This section describes how we extract $T_\mathrm{el}$ from the raw noise data seen in Figure \ref{supp_fig:raw_noise_1kOhm}.
Naturally, the cryoamp comes with its own input voltage and input current noise density figures, $e_\mathrm{n}$ and $i_\mathrm{n}$, respectively. The input voltage noise density is picked up by the amplifier directly, whereas the input current noise density causes a voltage drop $i_\mathrm{n} R$ across the sample that is yet again picked up by the cryoamp. The total noise therefore reads
\begin{align}
    S = \left(e_\mathrm{n}^2 + (i_\mathrm{n} R)^2 + J^2\right) G^2,
    \label{eqn:total_noise_ideal}
\end{align}
where $G$ is the cryoamp's (in our region of interest) frequency-independent gain and $J(R,T)$ is the sample's Johnson noise. 

However, having a closer look at the setup in Figure \ref{fig:exp_setup_resistance} reveals another component that adds to Equation \eqref{eqn:total_noise_ideal}: The resistance of the sample itself and the parasitic capacitance to ground of the wires between the cryoamp and the sample form a low-pass filter. This comprises the wire from the sample to the cryoamp and the leads used for the resistance measurement. Since they are connected to the sample in parallel, the individual capacitance of each wire can be summed up to the total capacitance that goes into the low-pass filter. This is the reason we measured the resistance in a 3-probe configuration. Adding another wire on the cryoamp's side of the sample would have further decreased the cutoff frequency of the low-pass filter. Typical values of our fridge for sample resistance and total wire capacitance are $R=\SI{1}{\kilo\ohm}$ and $C=\SI{600}{\pF}$, respectively, yielding a cutoff frequency of approximately $f_\mathrm{c}=\SI{265}{\kHz}$ \footnote{The formation of this natural low-pass filter is the reason this measurement was previously assumed unfeasible.}.

The low-pass filter affects all noise components of Equation \eqref{eqn:total_noise_ideal} propagating from the sample to the cryoamp; that is, all components except for the cryoamp's input voltage noise density. Taking into account the low-pass filter
\begin{align}
    L\left(f,R,C\right) = \frac{V_\mathrm{out}}{V_\mathrm{in}} = \frac{X_{\mathrm{C}}}{\sqrt{R^2 + X_\mathrm{C}^2}},
    \label{eqn:low-pass_filter}
\end{align}
where $X_\mathrm{C} = \frac{1}{2 \pi f C}$ is the reactance, we can rewrite Equation \eqref{eqn:total_noise_ideal} as
\begin{align}
    S^2 &= \left(e_\mathrm{n}^2 + (i_\mathrm{n} R L)^2 + (JL)^2\right) G^2 \nonumber\\
    &= (e_\mathrm{n} G)^2 + (i_\mathrm{n} R L G)^2 + (JLG)^2.
\end{align}
Redefining $LG$ as a frequency-dependent effective gain $G_\mathrm{eff}(f,R,C)$  finally leads to
\begin{align}
    S^2 = (e_\mathrm{n} G)^2 + \left((i_\mathrm{n} R)^2 + J^2\right)G_\mathrm{eff}^2.
    \label{eqn:total_noise_real}
\end{align}
Using this equation, we can calibrate the cryoamp on a run-by-run basis as follows: Firstly, we tune the sample into a state $T_\mathrm{ph} \gg T^\ast$---this simply means that $T_\mathrm{el} = T_\mathrm{ph}$---and use the magnetic field to tune the resistance to $\SI{1.29}{\kilo\ohm}$. We then measure the noise $S(T_1,\SI{1.29}{\kilo\ohm}) = S_1$. Secondly, we increase $T_\mathrm{ph}$ and decrease $B$ such that $R$ remains at $\SI{1.29}{\kilo\ohm}$, yielding a different noise value $S(T_2,\SI{1.29}{\kilo\ohm}) = S_2$. Subtracting $S_1$ from $S_2$ yields
\begin{align}
    S_2 - S_1 &= (e_\mathrm{n} G)^2 + \left((i_\mathrm{n} R)^2 + J_2^2\right)G_\mathrm{eff}^2 \nonumber\\
    &\hphantom{=} - \left((e_\mathrm{n} G)^2 + \left((i_\mathrm{n} R)^2 + J_1^2\right)G_\mathrm{eff}^2\right)    \nonumber\\
    &= \left(J_2^2 - J_1^2\right) G_\mathrm{eff}^2  \nonumber\\
    &= 4kR(T_2 - T_1) G_\mathrm{eff}^2,
    \label{eqn:calibration}
\end{align}
since all changes in the total noise can be attributed to changes in Johnson noise because the resistance is kept constant for both measurements. It is important that $T_\mathrm{el} = T_\mathrm{ph}$ when the calibration is done since we rely on a resistance thermometer, which measures $T_\mathrm{ph}$, to determine $T_1$ and $T_2$. Rearranging Equation \eqref{eqn:calibration} to
\begin{align}
    G_\mathrm{eff} = \sqrt{\frac{S_2 - S_1}{4 k R (T_2 - T_1)}} = L G = G \frac{X_{\mathrm{C}}}{\sqrt{R^2 + X_\mathrm{C}^2}}
    \label{eqn:effective_gain}
\end{align}
allows us to impose a physical function on $G_\mathrm{eff}$; namely that of a low-pass filter where the cryoamp's gain $G$ acts as a scale.

\begin{figure}[H]
    \centering
    \includegraphics[width=\columnwidth]{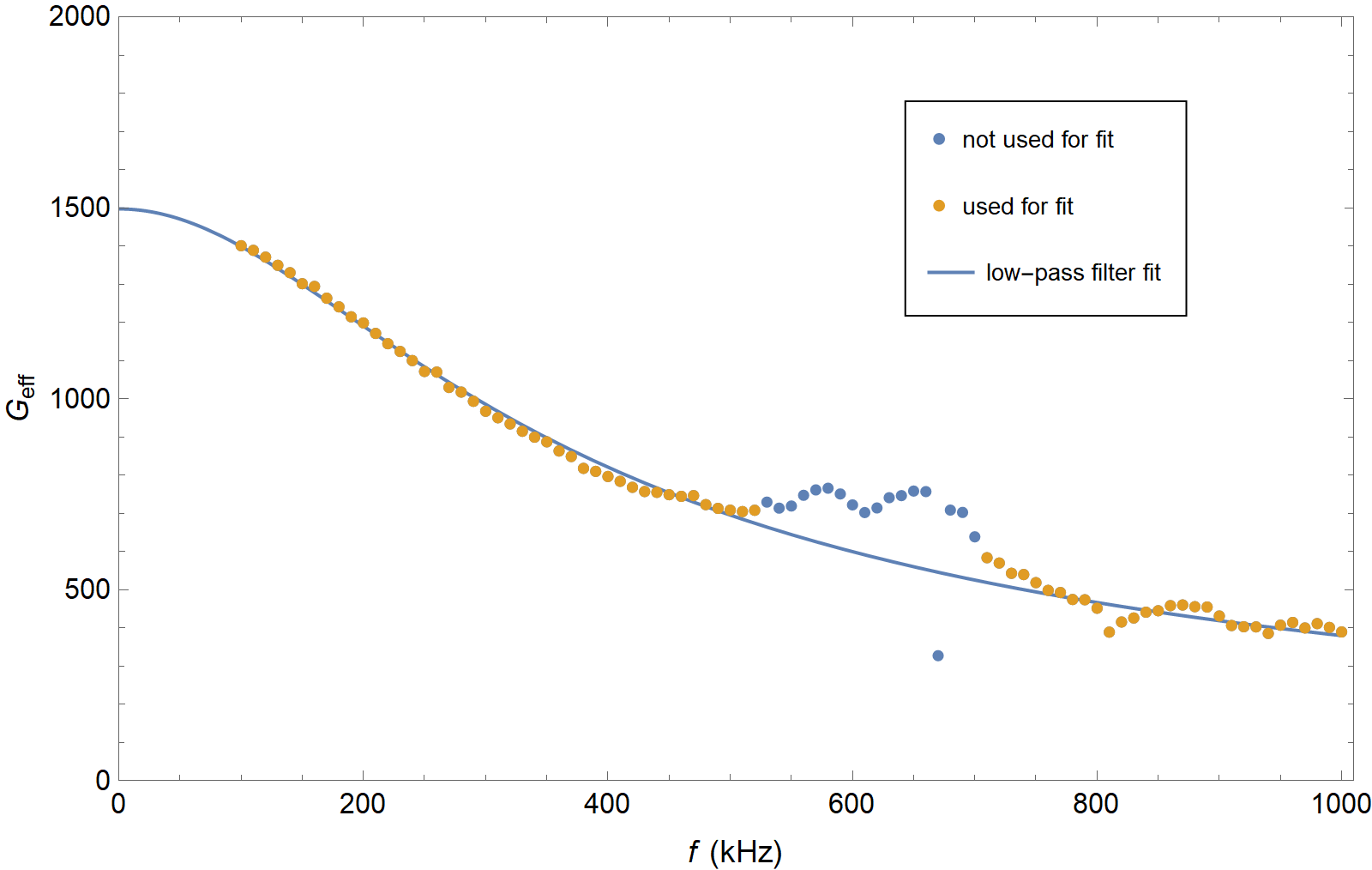}
    \caption{Resulting effective gain $G_\mathrm{eff}$ after calibration using data at $\SI{830}{\milli\kelvin}$ and $\SI{1820}{\milli\kelvin}$ with a fit according to Equation \ref{eqn:effective_gain}. The cutoff frequency is $f_\mathrm{c} = \SI{262}{\kilo\hertz}$}
    \label{supp_fig:lowpass_fit_1k}
\end{figure}

In the case of the $\SI{1.29}{\kilo\ohm}$ data, we used data at $T_\mathrm{el} = T_\mathrm{ph} = \SI{1820}{\milli\kelvin}$ and $\SI{830}{\milli\kelvin}$ for calibration. The result is shown in Figure \ref{supp_fig:lowpass_fit_1k}. We chose to exclude the data between $\SI{525}{\kilo\hertz}$ and $\SI{705}{\kilo\hertz}$ because the raw data are very noisy in this region.

Now that we know the effective gain at $\SI{1.29}{\kilo\ohm}$, we can subtract the Johnson noise from the data at $\SI{830}{\milli\kelvin}$, leaving only contributions from input voltage noise and input current noise. Then we subtract these contributions from all the other datasets, and divide by the low-pass filter fit we obtained in Figure \ref{supp_fig:lowpass_fit_1k} to obtain the Johnson noise before it is affected by the low-pass. The result is shown in the main body (Figure \ref{fig:results}b).

\section{3 kOhm data}
In order to verify our results, we repeated the noise measurement at a constant resistance of $R=\SI{3.00}{\kilo\ohm}$. The raw data are shown below in Figure \ref{supp_fig:raw_noise_3kOhm}.
\begin{figure}[H]
    \centering
    \includegraphics[width=\columnwidth]{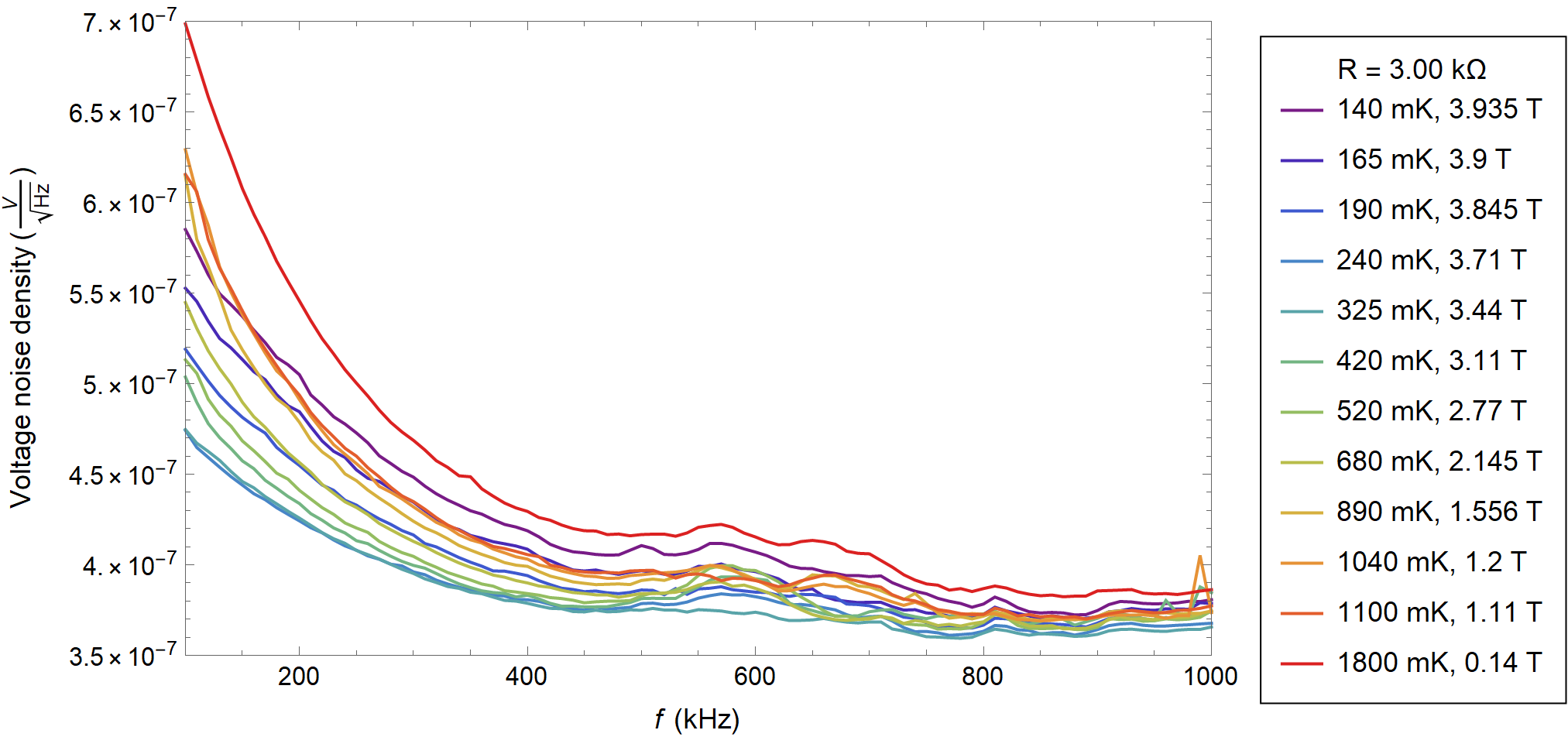}
    \caption{Voltage noise density as measured with our HF2LI LIA for $R = \SI{3.00}{\kilo\ohm}$. Data have been taken at 91 equidistant frequencies from $\SI{100}{\kilo\hertz}$ to $\SI{1}{\mega\hertz}$ with $\num{5E6}$ samples per frequency, a fourth-order low-pass filter and a time constant of $T_\mathrm{c}=\SI{7.812}{\micro\second}$ (BW NEP $\Delta f = \SI{10}{\kilo\hertz}$).}
    \label{supp_fig:raw_noise_3kOhm}
\end{figure}
We approach the detection limit of our measurement  as the increased resistance decreases the bandwidth of the noise measurement. For analysis of the $\SI{3.00}{\kilo\ohm}$ data we only considered frequencies $f < \SI{400}{\kilo\hertz}$ because of the noisy region centered around $\SI{500}{\kilo\hertz}$. Additionally, the low-pass filter effect greatly attenuates the signal above $\SI{600}{\kilo\hertz}$.

\begin{figure*}
    \begin{minipage}[c]{.49\textwidth}
        \begin{flushleft}
            \textbf{a}
        \end{flushleft}
        \centering
        \includegraphics[width=\columnwidth]{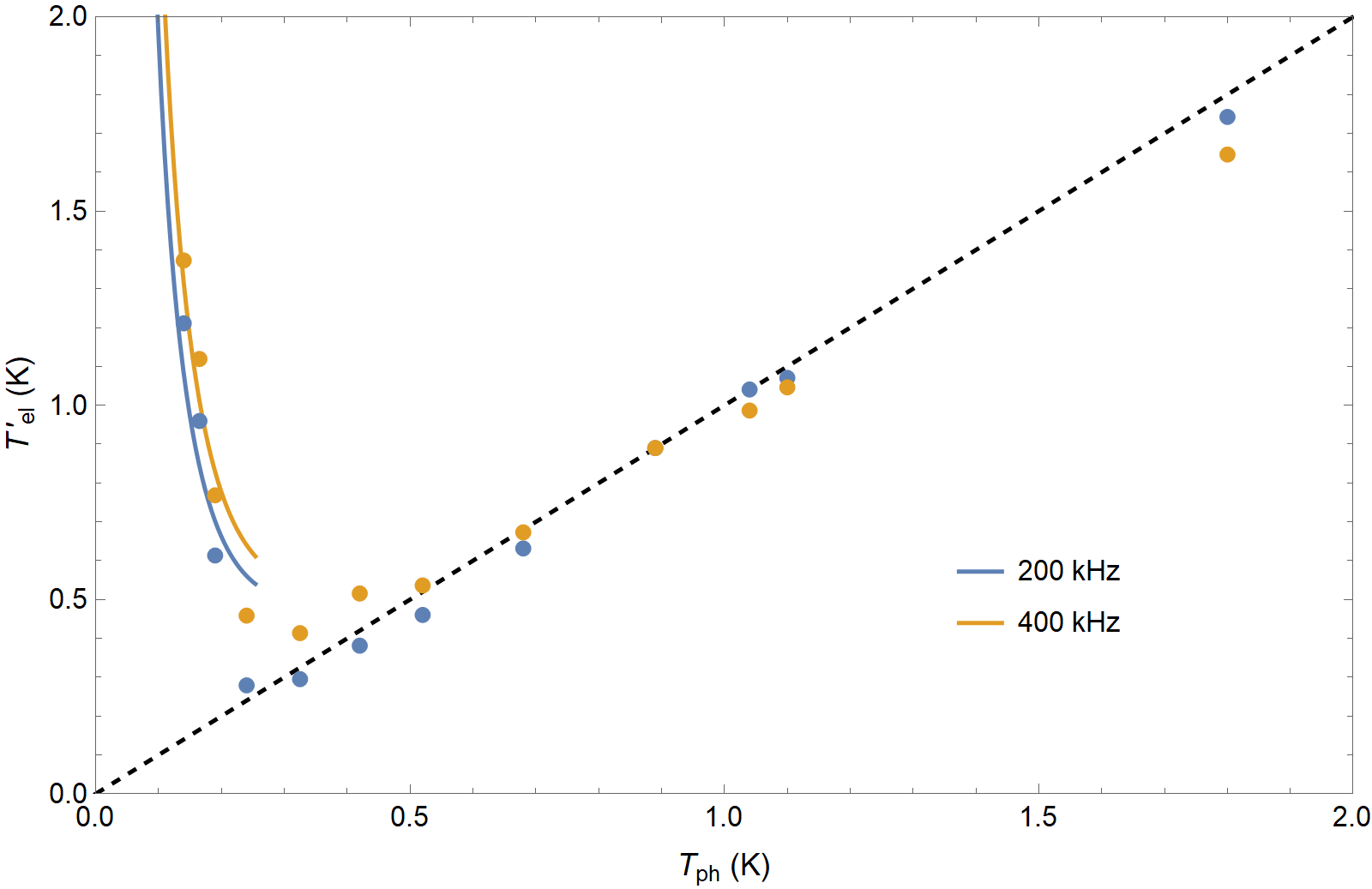}
    \end{minipage}
    \begin{minipage}[c]{.49\textwidth}
        \begin{flushleft}
            \textbf{b}
        \end{flushleft}
        \centering
        \includegraphics[width=\columnwidth]{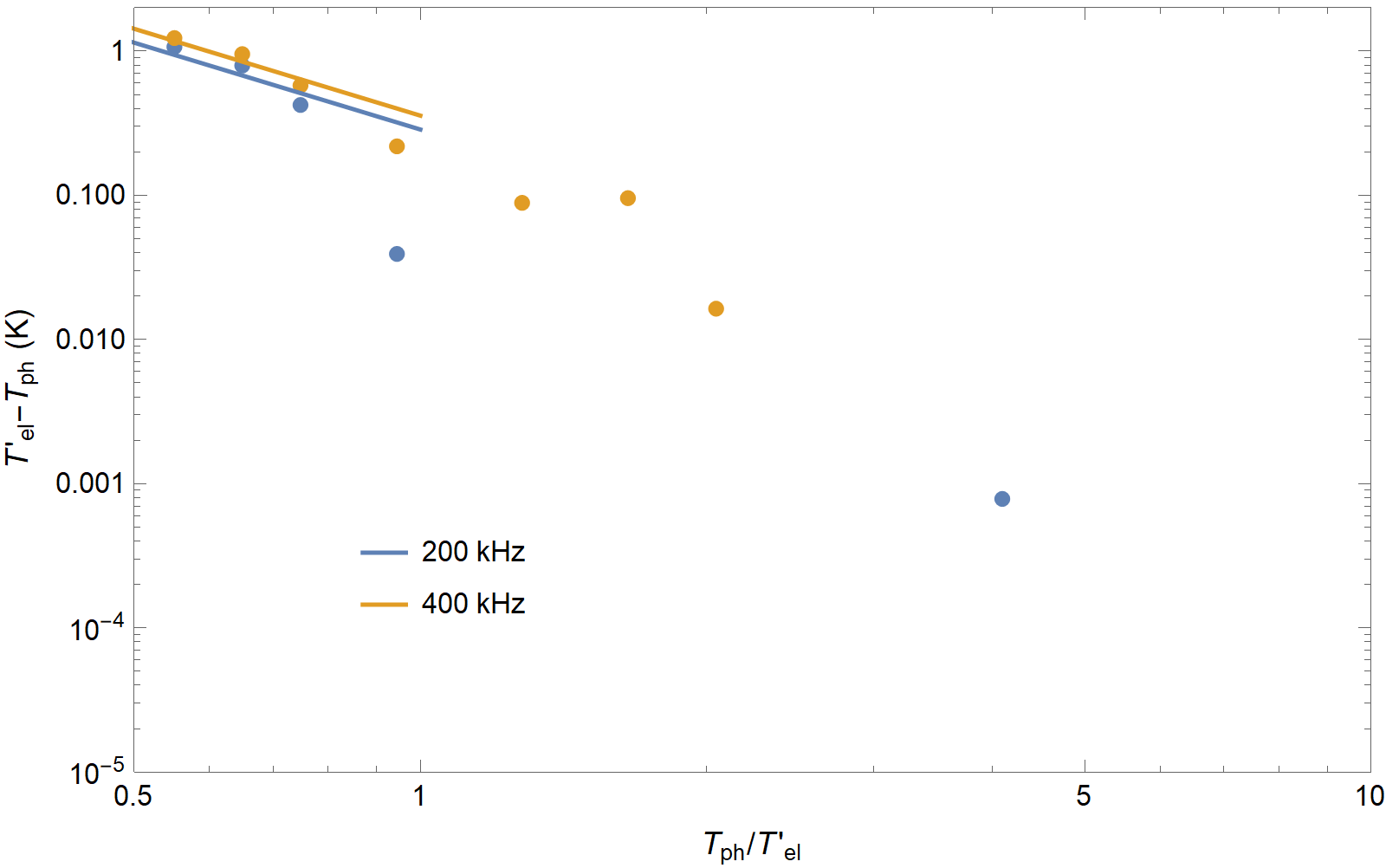}
    \end{minipage}
    \caption{\textbf{a} Ostensible electron temperature $T_\mathrm{el}^\prime = S_\mathrm{J}/4 k_\mathrm{B} R$ as a function of $T_\mathrm{ph}$ for $R = \SI{3.00}{\kilo\ohm}$. The dashed line marks $T_\mathrm{el}^\prime = T_\mathrm{ph}$. For $T_\mathrm{ph} > T^\ast = \SI{254}{\milli\kelvin}$, $T_\mathrm{el}^\prime$ and $T_\mathrm{ph}$ are in good agreement. We therefore identify it as the true electron temperature $T_\mathrm{el}$ in this regime. For $T_\mathrm{ph} < T^\ast$, $T_\mathrm{el}^\prime$ increases. Solid lines are fits $\propto T_\mathrm{ph}^{-2}$. \textbf{b} Difference of $T_\mathrm{el}^\prime$ and $T_\mathrm{ph}$ as a function of $T_\mathrm{ph}$ renormalized by $T^\ast$. As $T_\mathrm{ph}/T^\ast$ approaches unity, $T_\mathrm{el}^\prime$ starts to increase.}
    \label{supp_fig:tempComp3k}
\end{figure*}

Applying the same procedure as described above with $T_1 = \SI{890}{\milli\kelvin}$ and $T_2 = \SI{1800}{\milli\kelvin}$ yields the results presented in Figure \ref{supp_fig:tempComp3k}.
\end{document}